# Wafer-Scale Films of Two-Dimensional Materials via Roll-to-Roll Mechanical Exfoliation

Yigit Sozen[1]*, Thomas Pucher[1], Bhagyanath Paliyottil Kesavan[2], Nuria Jiménez-Arévalo[1], Julia Hernandez-Ruiz[1], Zdenek Sofer[2], Carmen Munuera[1], Juan J. Riquelme[1], and Andres Castellanos-Gomez[1]*

[1]*2D Foundry Research Group. Instituto de Ciencia de Materiales de Madrid (ICMM-CSIC), Madrid, E-28049, Spain.*

[2]*Department of Inorganic Chemistry, University of Chemistry and Technology Prague, Technická 5, 166 28 Prague 6, Czech Republic.*

*corresponding authors: yigit.sozen@csic.es, andres.castellanos@csic.es

## ABSTRACT

In this study, we demonstrate an improved version of the roll-to-roll mechanical exfoliation method, incorporating a controlled sliding motion into the exfoliation process to achieve uniform nanosheet films of two-dimensional materials at wafer-scale. This scalable technique enables the fabrication of high-quality films suitable for electronic and optoelectronic applications. We validate the process by fabricating $WSe_2$ phototransistors directly on the exfoliated films, achieving performance metrics comparable to the best reported devices based on electrochemically exfoliated material. The all-dry transfer method employed ensures minimal contamination and preserves the intrinsic properties of the material. This work highlights the potential of high-throughput mechanical exfoliation as a cost-effective and reliable route for large-scale production of 2D material-based devices.

**INTRODUCTION**

The exploration of two-dimensional (2D) semiconductor materials, particularly transition metal dichalcogenides (TMDs) like molybdenum disulfide ($MoS_2$) or tungsten diselenide ($WSe_2$), has received significant attention due to their exceptional electronic properties, such as high carrier mobility[1,2] and tunable band gaps[3–5], making them ideal candidates for applications in next-generation electronics, such as transistors and photodetectors[6–9]. Since the discovery of graphene in 2004[10], significant efforts have been made to develop scalable techniques for producing large-area films of 2D materials[11]. However, achieving wafer-scale production with low-cost techniques while maintaining high-quality material properties remains a significant challenge[12,13].

Recently, three main approaches have emerged for the scalable synthesis of 2D materials: liquid-phase exfoliation (LPE)[14,15], electrochemical exfoliation (EE)[16], and chemical vapor deposition (CVD)[17,18]. Among these, LPE and EE have provided low-cost, scalable routes for producing solution-processed 2D material-based electronics. However, LPE-based devices typically suffer from low carrier mobility (often < 1 cm²/V·s) and modest $I_{on}/I_{off}$ ratios, largely due to solvent residues, structural defects, and poor alignment of the exfoliated flakes, which degrade electronic performance[19–22]. Most previous LPE-based studies have gated the channel material with ionic liquids, which offer notable advantages such as capacitances that are orders of magnitude higher than standard oxide dielectrics and the ability to achieve large carrier density modulation at low gate voltages[23]. Nevertheless, these non-standard gating approaches introduce practical challenges, including slow switching speeds, increased fabrication complexity, and limited compatibility with conventional device architectures.

In contrast, EE methods generally produce flakes of higher quality compared to LPE, and some reports have demonstrated mobilities exceeding 10 cm² V$^{-1}$ s$^{-1}$[24–26]. However, these results are often achieved under optimized conditions involving carefully engineered chemical interfaces and/or doping strategies. EE-derived films are typically deposited using the Langmuir–Blodgett technique[27,28], which provides precise control over monolayer assembly. Despite this, the Langmuir–Blodgett method is limited in throughput and is not well suited for repeated multilayer deposition, restricting its practicality for rapid wafer-scale fabrication and flake density optimization.

CVD has proven to be an effective method for producing high-quality monolayers with excellent electronic properties, particularly for materials like $MoS_2$ and $WSe_2$[29–32]. Moreover, CVD enables the growth of continuous, wafer-scale films, which is crucial for integrating 2D materials into industrial-scale applications[11,33,34]. The mobility of $MoS_2$ and $WS_2$ produced by CVD can range from 0.2 to 70 cm² V$^{-1}$ s$^{-1}$, depending on growth conditions, which are also constrained by grain boundary formation and the inhomogeneity of crystallization seeds and precursor delivery.[35,36] Recently, Lee et al. introduced a scalable "roll-printing" technique capable of producing meter-scale van der Waals films[37]. This method enables the uniform deposition of various 2D materials onto a wide range of substrates through the use of a lubricant during the printing process. The resulting films, however, are relatively thick (~2 μm), which poses limitations for their use in flexible electronics. Moreover, electronic and optoelectronic performance of fabricated devices may not yet fully meet the requirements of advanced applications, highlighting the need of further optimizations.

In this context, our group has previously developed a high-throughput mechanical exfoliation method to produce large-area 2D material films[38]. While this approach enabled the production of high-quality films, the size of the exfoliated areas remained limited to the ~ 13 cm$^2$ scale. In this work, we present an enhanced version of our mechanical exfoliation setup with a roll-to-roll-like geometry that involves laterally sliding one of the rolls during the exfoliation process. This modification significantly improves the homogeneity of the produced films, allowing us to achieve uniform, large-area films with sizes up to ~ 5.4 inch² (~ 35 cm$^2$). Using this method, we demonstrate the fabrication of homogeneous films of $WSe_2$, $MoS_2$, hexagonal boron nitride (h-BN) and graphite on 2-inch wafers. As a representative example, $WSe_2$ was selected for device fabrication, since systematic benchmarking across multiple materials at large scale is experimentally impractical. From these films, we fabricated phototransistors, achieving mobility values that surpass those reported for devices fabricated with LPE[19–22], and are within the same order of magnitude as FETs built from electrochemically exfoliated flakes[24–26,39,40].

The high-throughput mechanical exfoliation method presented here offers significant advantages over both LPE and CVD. Unlike LPE, it eliminates the issues related to solvent residues and misalignment of flakes, producing films with higher homogeneity, lower porosity and fewer defects. It also avoids the grain boundary issues typical of CVD-grown films, as the exfoliated flakes preserve their crystallinity and form interfaces with reduced structural disorder, leading to improved charge transport properties. Moreover, the method is cost-effective and scalable, offering a practical route for the large-area production of high-quality 2D materials.

Despite the advantages, one limitation of this approach is the reduced control over film thickness at the atomic scale, since the films are composed of randomly oriented multilayered flakes rather than strictly monolayer assemblies. Nonetheless, the resulting films (typically a few tens of nanometers thick) remain well-suited for a wide range of electronic and optoelectronic applications where thin, continuous, and high-mobility layers are required. Overall, our results highlight the potential of this method for the scalable fabrication of 2D material-based electronic devices and its relevance for future integration into next-generation technologies.

**RESULTS AND DISCUSSION**

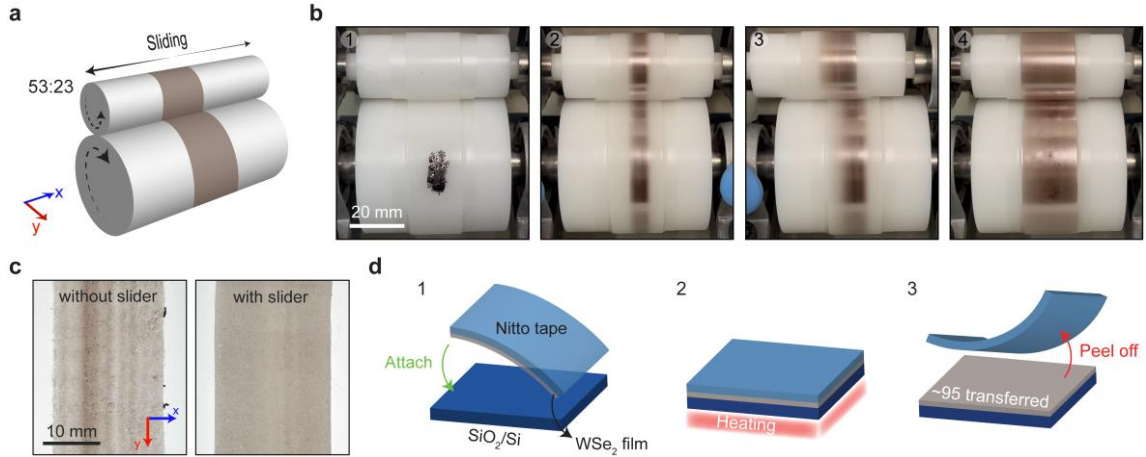

**Figure 1. Roll-to-roll exfoliation of WSe$_2$ crystal through the semi-automated setup including a horizontal slider.** (a) Schematic illustration of the roll-to-roll exfoliation setup emphasizing the integration of linear sliding feature which improves uniformity of two-dimensional films. (b) Sequential images captured at various phases of the exfoliation process of the WSe$_2$ crystal: from bulk crystal loading to the formation of a thin film. (1) Image showing two in-contact parallel cylinders covered with Nitto tape, one of which is loaded with a bulk WSe$_2$ crystal. (2) Image captured immediately after the initiation of the exfoliation process. (3) Sliding of the top cylinder improves the spatial distribution of exfoliated WSe$_2$ flakes across the whole tape. (4) Final state of the tapes after the exfoliation process. (c) As-fabricated WSe$_2$ nanosheet films on Nitto tape obtained from the exfoliations without (left) and with (right) the use of the sliding feature. (d) Schematic describing the thermal release process of exfoliated films from Nitto to the target substrate. (1) The Nitto tape containing the exfoliated film is attached onto target substrate, (2) heated at 100 °C for 5 minutes, and (3) peeled off, yielding a 95% transfer of the thin film onto the substrate.

Figure 1a provides a simplified illustration of the adjustable rolling exfoliation mechanism, including two cylinders with a diameter ratio of 53:23. In this study, unlike the initial setup presented in our previous paper, the roll-to-roll setup was modified by integrating a linear bearing slide. With this modification we can laterally displace one of the cylinders during the exfoliation while maintaining continuous contact. As a result of selecting prime numbers for the diameter ratio, the commensurability between the two cylinders is reduced. This ensures that the cylinders will only align at the same initial contact point again after 1219 revolutions. The extended time interval before realignment promotes a more even distribution of flakes along the exfoliation direction, which is defined as y-axis in Figure 1a. On the other hand, the integration of the sliding mechanism prevents the formation of fringe patterns along the y-axis, which are typically caused by thickness variations in the bulk crystal during loading. Basically, this modification enables the flakes to spread more evenly along the direction perpendicular to the exfoliation axis (x-axis in Figure 1a), thereby promoting more uniform film formation.

Figure 1b illustrates a series of captured images at different stages of the exfoliation of a bulk $WSe_2$ crystal, conducted with an adjustable linear motion mechanism. The first image shows the moment when the bulk $WSe_2$ crystal was loaded into the middle of the source Nitto tape attached to the cylinder with the larger diameter. The Nitto tape, with its adhesive side facing outward, was mounted onto the cylinders via pre-attached double-sided tape. The second image was captured immediately after the initiation of the exfoliation process by rotating two contacting cylinders with a motor, illustrating the exfoliation and transfer of flakes from the source tape to the receiving tape. The third image was taken after slowly sliding the top cylinder laterally. This lateral movement ensures an even distribution of the flakes, independently of the position and size of the original bulk $WSe_2$ crystal, across the entire tapes. The linear motion is repeatedly maintained back-and-forth during the exfoliation to achieve evenly coated films. The resulting tapes after exfoliation are shown in the fourth image, which reveals uniform film formation.

To assess the effect of the lateral sliding mechanism on film quality, exfoliations were carried out both with and without this feature. As shown in the left panel of Figure 1c, thickness fringe formations along the y-axis are more pronounced when the sliding function is omitted, which originates from the non-uniform distribution of the bulk crystal on the source tape. In contrast, when the sliding mechanism is engaged during exfoliation, the flakes are more evenly distributed across the entire tape, effectively suppressing fringe formation and yielding more uniform films, as shown in the right panel of Figure 1c.

Figure 1d illustrates the thermal release process used to transfer two-dimensional films onto a target substrate. Following the exfoliation, the resulting tape is attached to the target substrate by gently pressing with a cotton swab and then heated at 100 °C for 5 minutes to induce the thermal release of the exfoliated nanosheets. After heating, the tape is slowly peeled off with tweezers, resulting in approximately 95% transfer efficiency of the nanosheets from the Nitto tape to the substrate.

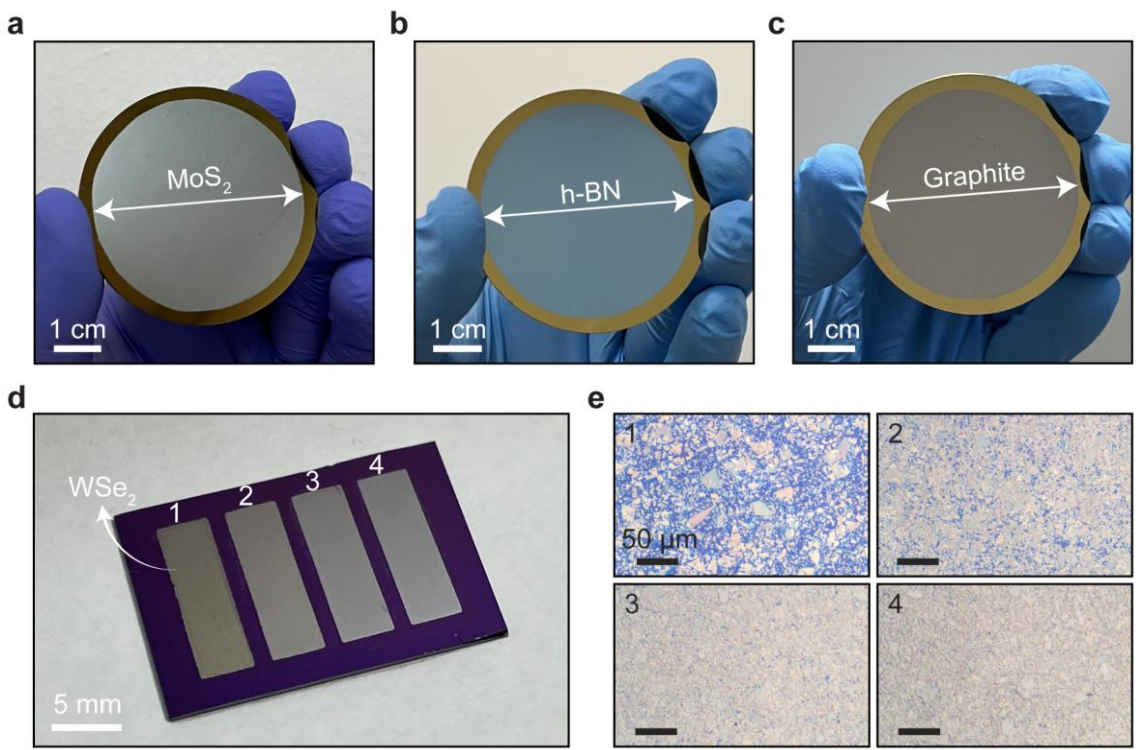

**Figure 2. Large area films of 2D crystals on 2-inch SiO$_2$/Si wafers and effect of successive film deposition on continuous film formation.** Large area films of (a) MoS$_2$, (b) h-BN, and (c) graphite on SiO$_2$/Si wafers. (d) WSe$_2$ films on SiO$_2$/Si substrate consist of 1, 2, 3, and 4 sequential transfers. (e) Reflection-mode micrographs acquired from the films represented in (d) indicating the formation of a continuous film of interconnected flakes with increasing number of transfers.

As the next step, we focused on the production of roll-to-roll exfoliated 2D materials on wafer-scale SiO$_2$/Si substrates to demonstrate the scalability of the method and its compatibility with industrial applications. To perform large-area film exfoliation, it is only required to scale up the size of the double-sided tape and the Nitto tape. We address the reader to Video S1 to show the fabrication of large-area $MoS_2$ films. First, bulk $MoS_2$ flakes were distributed onto the source tape along a common axis, and then exfoliation was carried out until a homogeneous film formation is observed on the receiving tape, which usually takes about a minute. As a representation, Figure S1 provides an image taken from the large-area $WSe_2$ film on Nitto tape after exfoliation. The large area films were then transferred onto inch-sized $SiO_2$/Si wafers by following the same thermal releasing process explained in the previous section. This process was repeated successively to obtain high coverage on the substrate (see further explanation in the paragraph below). Figure 2a-c show large area film formations for $MoS_2$, h-BN, and graphite on 2-inch $SiO_2$/Si wafers, after 4, 5, and 5 film transfers, respectively.

While a single transfer of exfoliated flakes can achieve relatively high substrate coverage, often approaching 80% (see Figure S2), the resulting film typically consists of isolated or loosely connected flakes. For applications requiring efficient charge transport or optical uniformity, such as in electronics and optoelectronics, a continuous film with a well-connected flake network is essential. This type of film ensures good percolation pathways and improved overall performance. To achieve this, multiple deposition cycles can be performed by sequentially stacking several films. Each additional layer increases the coverage and significantly enhances the interconnection between flakes, leading to a more uniform, continuous, and structurally reliable thin film. To experimentally demonstrate the impact of multiple transfers on uniform film formation, we prepared multiple $WSe_2$ films on a $SiO_2$/Si substrate with transfer numbers varying from 1 to 4 (see Figure 2d). Reflection-mode optical micrographs represented in Figure 2e demonstrate enhanced coverage and film continuity on the target substrate, as the number of transfer cycles increases.

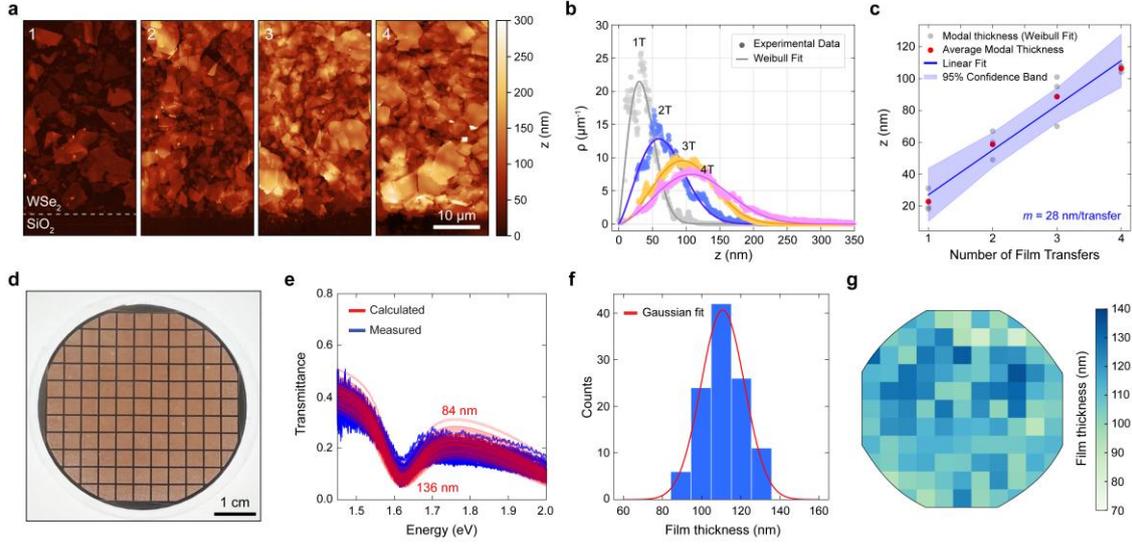

**Figure 3. AFM characterization of WSe$_2$ film thickness evolution with successive transfers, and large-area thickness estimation on polycarbonate (PC) using an optical model based on the Beer-Lambert law.** (a) AFM topography images taken from WSe$_2$ films on SiO$_2$/Si substrate, consisting of 1, 2, 3, and 4 transfers. Images highlight that subsequent film transfer results in well-covered film formation with the stacking of each additional layer. (b) The evolution of the height distribution in films with increasing number of film transfers. Solid lines correspond to Weibull distribution fits for each experimental dataset, which are performed to determine the modal thickness of each film. (c) The change in the modal thickness of thin films as a function of the transfer number, giving rise to a thickness of 28 nm per transfer. Linear regression line and 95% confidence band in the plot were obtained after averaging the estimated modal height values extracted from 3 distinct regions for each transfer step. (d) Large-area WSe$_2$ film formation on PC after 4 transfers, where an aluminum grid mask was placed on top in order to perform spatially resolved transmission spectrum measurements. (e) Experimentally obtained transmittance spectra from grids with the corresponding calculated spectra through the Beer-Lambert law. (f) Histogram showing the distribution of estimated film thickness over the large-area WSe$_2$ film. Gaussian distribution curve (red line) gives an approximate mean value of 111 nm. (g) Heat map showing the spatial variation of the estimated thickness across the wafer.

To investigate how successive transfers affect film thickness and uniformity, we performed Atomic Force Microscopy (AFM) measurements on WSe$_2$ films deposited on SiO$_2$/Si substrate. Additionally, this characterization is necessary to assess the optimal transfer number required for continuous film formation. Prior to AFM characterization, 5-µm-wide scratches were made on the film using a vinyl record player stylus to expose the underlying substrate, allowing for accurate measurement of the height difference between the substrate and the transferred films. AFM imaging was acquired using a scan area of 50 × 50 µm$^2$.

Figure 3a shows selected portions of the AFM topography images taken from $WSe_2$ films consisting of 1, 2, 3, and 4 transfers, displayed side by side to facilitate direct comparison. Full scan images are represented in Figure S3. The AFM image taken from the first transfer shows that flakes are discontinuous and only partially overlapping, leaving some uncovered areas. After the second transfer, flakes start to settle in uncovered regions, increasing the flake-to-flake overlap, and start to form continuous pathways. By stacking more flakes after third and fourth transfers, we ensure near full coverage, and begin to obtain a continuous film owing to the formation of a highly interconnected flake network formation.

The average film thickness also increases with the transfer of each additional layer, as evidenced by the gradually increasing brightness in the topography images. To quantitatively investigate the evolution of the film thickness, we obtained histogram plots showing the height distribution of each transfer step, which are represented in Figure 3b. In order to accurately determine the most probable thickness values, we quantify the modal thickness for each transfer step by performing a Weibull fit to the height distributions. The modal thickness of the films was extracted numerically from the fitted curve data. From the corresponding datasets of the films with 1, 2, 3, and 4 transfers, the modal thicknesses were estimated as 31, 60, 95, and 108 nm, respectively.

Each additional transfer not only increases the film thickness but also broadens the distribution. This broadening with additional transfers does not indicate any imperfection in the transfer process; instead, it reflects the statistical nature of independently stacked layers. With each transfer, the thickness of the film increases with the sum of independent Weibull-distributed layers. As a result, the cumulative thickness distribution after $n$ transfers can still be well-approximated by a Weibull fit with identical shape ($k$) and with a larger scale parameter that scales as $\lambda_n = \lambda n^{1/k}$.

The observed broadening is therefore a direct consequence of the increase in both the mode and the variance with each transfer, and fully consistent with the additive nature of independent stochastic layers.

To conduct a more accurate analysis for determining the trend in thickness increase per transfer step, we performed 3 different AFM scans at different locations for each studied sample. After extracting the modal thicknesses of each dataset, we calculated the average modal thickness for each transfer step and performed linear regression to establish a clearer relationship between the number of transfer steps and film thickness increase. Figure 3c shows determined modal heights and their averages as a function of the number of film transfers along with the corresponding linear regression fit. Accordingly, the slope of the linear fitting line, which represents the thickness increase per transfer, was obtained as 28 nm/transfer. These AFM results collectively indicate that each transfer contributes an additional layer of material, confirming that the process effectively stacks one film onto another without a significant loss in deposition efficiency at successive steps.

Moreover, we determined the length and height distribution of individual $WSe_2$ flakes using AFM images acquired from a single film transfer onto a $SiO_2$/Si substrate. In total, 139 nanosheets were taken into account and the statistical output is represented in Figure S4. According to the lognormal fit, the mean values for the length and the thickness of the flakes were obtained as 3.4 μm and 24 nm, respectively, corresponding to a length-to-thickness aspect ratio of ≈142, which is higher than roll-to-roll $MoS_2$ nanosheets reported in our previous study[38] and other nanosheet flakes fabricated with LPE[41,42] and EE methods[39,43,44]. This high aspect ratio is advantageous for forming percolating conductive networks, as it promotes interflake connectivity at lower loading densities and enhances charge transport across the film[44].

Although AFM is a highly precise technique to study the local thickness and topography of the films, it is impractical for mapping the thickness over a 2-inch-sized film as it would be slow and time-consuming. To estimate the spatial variation in thickness across an inch-sized $WSe_2$ film transferred onto a polycarbonate (PC) wafer, we developed a Beer-Lambert law-based optical model using the reported refractive index for multilayer $WSe_2$[45]. First, we prepared a large-area continuous $WSe_2$ film on a transparent PC substrate by performing 4 film transfers. Afterwards, to enable spatial thickness mapping, we placed an aluminium grid mask on top, dividing the film into smaller regions (see Figure 3d). We then collected the transmitted light intensity from each grid by using a 10× objective lens (NA = 0.3). The details of the experimental setup for the transmittance spectroscopy can be found in our previous study[46]. The transmittance spectrum of the films ($T_{film}$) was determined by normalizing the transmitted light intensity measured through the film/substrate stack ($I_{film+substrate}$) to that of the bare substrate ($I_{substrate}$) as given by the following expression:

$$T_{film} = \frac{I_{film+substrate}}{I_{substrate}}$$

The Beer-Lambert law quantitatively relates the transmittance spectrum of a material to it's absorption coefficient and thickness. In terms of transmission, the Beer-Lambert law can be written as

$$T = e^{-(\alpha+f)d}$$

where $\alpha$ is the absorption coefficient, $d$ is the film thickness, and $f$ is the scattering factor, which is an essential parameter to account for diffuse scattering in the film. $\alpha$ can be expressed in terms of the extinction coefficient ($k$), and the equation becomes

$$T = e^{-(\frac{4\pi k}{\lambda}+f)d}$$

By using this model, we estimated the full transmittance spectra of $WSe_2$ as a function of thickness for different empirical scattering factors. We selected the scattering factor that approximately yielded the best agreement between the experimental and simulated curves. To completely reproduce the experimental data, we applied a spectral red shift of approximately 10 nm and smoothed the calculated spectra using a 20 nm moving average. This shift and the observed broadening of the A exciton feature in the measured optical data were attributed to biaxial strain introduced during thermal transfer, resulting from the mismatch in thermal expansion coefficients between $WSe_2$ and the PC substrate. Finally, the experimental transmittance values at the A exciton energy were compared to the ones in calculated curves to estimate the local thickness at each measured point.

Figure 3e gives all the experimental and corresponding calculated spectra. Accordingly, the majority of the experimental spectra fall within a thickness range of 84 nm to 136 nm. The histogram in Figure 3f shows the distribution of the estimated thicknesses. By performing Gaussian curve fitting, we obtained a mean of 111 nm, which agrees well with the thickness value measured with AFM. Based on the estimated thickness values, we generated a heat map to show the spatial distribution of thickness across the wafer (see Figure 3g). It should be noted that this method does not yield absolute thickness values without prior calibration of the scattering factor, which must be manually adjusted to match the measured spectra. As such, some uncertainty in the absolute thickness is expected. However, despite this limitation, the model reliably captures relative thickness variations across the wafer, providing valuable insight into the film's spatial homogeneity. While simplified, the approach is sufficiently robust for assessing uniformity, which is a critical parameter in wafer-scale film fabrication.

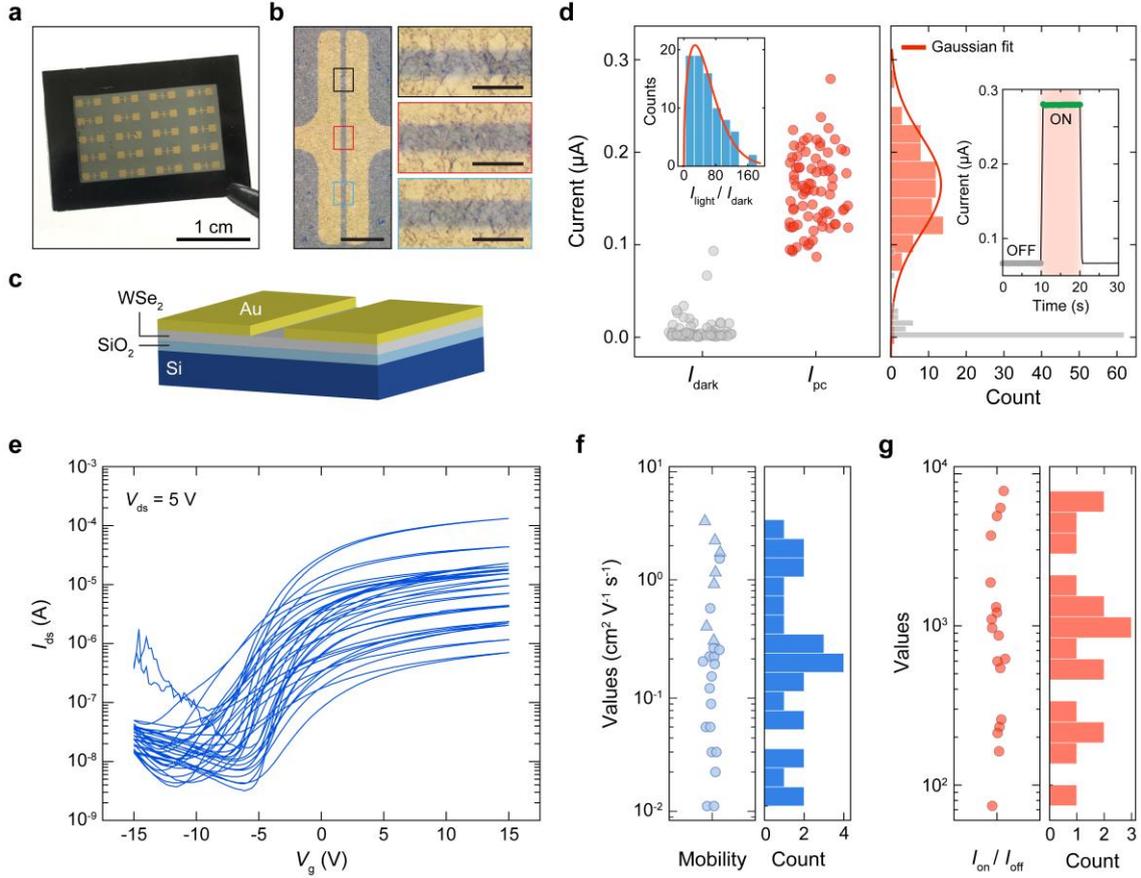

**Figure 4. Large batch fabrication and characterization of WSe$_2$ phototransistors.** (a) Optical image of a single batch containing 20 WSe$_2$ devices. (b) Optical microscope image showing the entire channel region of a representative device (left), along with magnified views of three distinct locations within the channel (right). The scale bars correspond to 200 μm (left) and 50 μm (right). (c) Three-dimensional schematic illustration of a representative device structure. (d) Dark current ($I_{dark}$) and photocurrent ($I_{pc}$) values collected from 79 WSe$_2$ devices, with corresponding histogram plots illustrating their distribution. A Gaussian fit to the photocurrent distribution data yields a mean of 0.16 μA. The inset on the left panel shows the distribution of the ratio between the current measured under illumination ($I_{light}$) and in the dark ($I_{dark}$), $I_{light} / I_{dark}$. The inset on the right panel presents a time-resolved photocurrent response from one of the devices, where OFF and ON state data are highlighted in grey and green, respectively. (e) Transfer characteristics of 18 back-gated WSe$_2$ transistors fabricated on a Si substrate with 90 nm SiO$_2$ layer. (f) Calculated mobility ($\mu$) values for 25 WSe$_2$ transistors with a histogram showing the distribution. Circles indicate the data extracted from the transfer curves given in (e), whereas the triangles represent the data of additional devices from different batches with varying channel lengths. (g) $I_{on}/I_{off}$ values extracted from (e).

To ensure a robust statistical evaluation of the electronic and optoelectronic performance, and to assess the reproducibility of devices fabricated by roll-to-roll films, we performed large-batch production of WSe$_2$ devices. In this process, we fabricated four device chips on silicon substrates with a 290 nm silicon oxide layer, each containing 20 individual devices (see Figure 4a), resulting in a total of 80 devices. For each chip, four sequential transfers of WSe$_2$ films were carried out, followed by the formation of source-drain contacts (30 μm channel length and 1 mm channel width) on top of the films via thermal evaporation of 80 nm of Au using shadow masks. Out of the 80 fabricated devices, 79 exhibited functional electronic behavior, corresponding to a device yield of 98.8%. Figure 4b presents optical microscope images of the entire channel region of one of the devices (left), with magnified images captured from three distinct locations along the channel (right). These zoomed-in images reveal that the whole channel is well-covered with interconnected $WSe_2$ flakes bridging the source and drain electrodes. Figure 4c shows a three-dimensional schematic illustration of the device structure, providing a visual representation of the device configuration.

For the operational devices, we performed time-resolved photocurrent measurements at a constant bias voltage of 5 V by switching from dark to illuminated state at every 10 s, using an excitation wavelength ($\lambda$) of 625 nm with an incident light power ($P_{light}$) of 0.6 mW. We extracted the mean values for current measured under dark ($I_{dark}$) and the current measured under illuminated state ($I_{light}$) by averaging the data recorded at each state, and then calculated the net photocurrent as $I_{pc} = I_{light} - I_{dark}$ (see inset in the right panel of Figure 4d). Left panel in Figure 4d shows all extracted $I_{dark}$ and $I_{pc}$ values from 79 $WSe_2$ devices. The majority of the devices exhibit low $I_{dark}$, on the order of $10^{-9}$ A. The calculated $I_{pc}$ values range from 0.09 μA to 0.37 μA, with a mean value of 0.16 μA, according to a Gaussian fit of the histogram shown in the right panel of Figure 4d. As represented in the inset of the right panel in Figure 4d, the devices exhibit square-shaped photocurrent cycles, which typically indicate fast and stable responses to light excitation. The device response times are less than 20 ms, however this value represents an upper bound, as the measurement was limited by the resolution of the experimental setup. The inset in the left panel of Figure 4d shows distribution of the ratio of $I_{light}$ to $I_{dark}$. According to a Weibull fit, the histogram yields a mode of 28; however, the distribution is right-skewed, indicating a considerable number of devices with much higher $I_{light}/I_{dark}$ ratios ranging from 60 to 130.

To collect statistical data on the FET characteristics of WSe$_2$ devices, a separate chip was fabricated on a thinner 90 nm SiO$_2$ layer, which offers improved gate control and reduced threshold voltage ($V_{th}$) compared to the conventional 290 nm oxide. The transfer characteristics of these devices were evaluated using back-gate modulation through the SiO$_2$ dielectric. After four sequential film transfers, source-drain electrodes with a channel length of 30 µm and a channel width of 1 mm were deposited via thermal evaporation (5 nm Cr and 45 nm Au) using a shadow mask. Transfer curve measurements were performed by sweeping the gate voltage ($V_g$) from -15 V to 15 V with a sweep rate of 0.1 V s$^{-1}$, under a drain-source voltage ($V_{ds}$) of 5 V. As seen in Figure 4e, the devices exhibit similar transfer characteristics, reflecting the uniformity and reproducibility of the roll-to-roll WSe$_2$ devices. All devices show weak ambipolar characteristics, with significantly stronger electron conduction (n-type) compared to the minor hole conduction (p-type), as evidenced by the higher drain-source current ($I_{ds}$) levels in the n-type regime. The transition point, corresponding to the $V_g$ at which the device shifts from n-type to p-type conduction, shows slight variation across devices.

Based on the set of transfer curves presented in Figure 4e, we extracted key FET performance metrics, such as mobility and I$_{on}$/I$_{off}$ ratios (see Figure 4f and 4g). We also included the mobility values obtained for the best performing devices, which were obtained from separate fabrication batches and are indicated by triangles in Figure 4f. The mobility ($\mu$) was extracted using the equation:

$$\mu = \frac{dI_{ds}}{dV_g} \frac{L}{W_{eff}} \frac{1}{C_G \cdot V_{ds}}$$

where $I_{ds}$, $V_{ds}$, and $V_g$, are the drain-source current, drain-source voltage, and gate voltage, $C_G$ is the capacitance of the silicon oxide dielectric, $L$ and $W_{eff}$ are the channel length and effective channel width, respectively. $W_{eff}$ can be estimated by multiplying the total channel width, $W$, with a factor $c$ that defines the percentage of the channel area covered by flakes (0 refers to completely empty channel and 1 refers to completely covered channel). Since the channels of the devices are mostly well-covered with flakes, and individually determining the coverage for each device channel is not practical, a uniform $c$ parameter of 0.9 was considered for the large-batch mobility calculations. Owing to the significant dominance of the n-type conduction in the devices, we estimated electron mobilities from the linear part of the n-type regime. The calculated mobility values span a broad range, from 0.01 to 3.4 cm$^2$ V$^{-1}$ s$^{-1}$, with 68% of the devices exhibiting mobility values above 0.1 cm$^2$ V$^{-1}$ s$^{-1}$. For each transfer curve presented in Figure 4e, we also calculated $I_{on}/I_{off}$ ratio. Since devices reach saturation point when $V_g$ was swept toward more positive voltages, $I_{on}$ was defined as the maximum $I_{ds}$ within the n-type branch of the transfer curve, while $I_{off}$ was taken as the $I_{ds}$ measured at the transition point. Accordingly, the calculated $I_{on}/I_{off}$ ratios range from $10^2$ to $10^4$ (see Figure 4g).

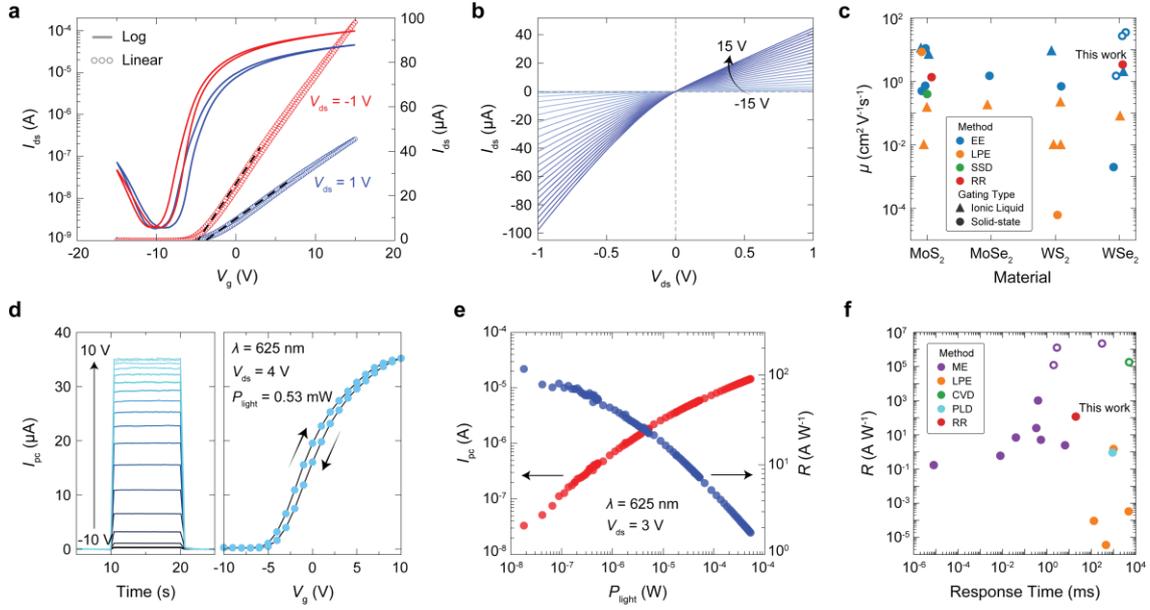

**Figure 5. Evaluating the WSe$_2$ devices as a phototransistor.** (a) Semilog and linear transfer curve characteristics of the back-gated WSe$_2$ phototransistor recorded at forward (1 V) and reverse (-1 V) $V_{ds}$ conditions. Blacked dashed lines correspond to linear extrapolations to extract threshold voltage ($V_{th}$). (b) *IV* curves with a $V_g$ sweep from -15 V to 15 V in steps of 1 V. (c) Comparison of mobility values reported for TMD-based devices obtained with other low-cost fabrication methods. (d) The evolution of single-cycle photocurrent response when $V_g$ is swept from -10 V to 10 V (left) and corresponding $I_{pc}$ values as a function of the $V_g$ (right). (e) Variation in $I_{pc}$ and responsivity ($R$) as a function of incident light power ($P_{light}$). (f) Comparison of $R$ versus response time characteristics for WSe$_2$-based devices fabricated by different methods. Open symbols in (c) and (f) are used to mark data points corresponding to studies that employed functionalization or device engineering techniques to improve device performance. Note that panels (a) and (b) present results from one device, whereas panels (d) and (e) correspond to measurements from a separate device.

To gain deeper insights, we evaluated the performance of representative high-performing WSe$_2$ devices through detailed electronic and optoelectronic measurements. Figure 5a gives the transfer characteristics of the device when it is operated both at forward and reverse $V_{ds}$ conditions. As reported in the previous section, it exhibits n-type conduction dominated ambipolar characteristics for a consecutive forward/reverse $V_g$ sweeping with a step size of 0.2 V s$^{-1}$. A narrow hysteresis is apparent within the entire gate range, which can be attributed to the absence of charge trapping inside the channel, interface trap states, and trap states originating from surface adsorbates[47]. By performing a linear extrapolation to the linearly scaled transfer curves, we extracted $V_{th}$ for forward and reverse $V_{ds}$ conditions, which are 3.7 V and 4.7 V, respectively. From the n-type regimes, electron mobilities were calculated to be 1.7 (3.4) cm$^2$ V$^{-1}$ s$^{-1}$ for forward (reverse) bias conditions. $I_{on}/I_{off}$ ratios were calculated to be above $10^4$, representing the highest value obtained among all fabricated WSe$_2$ devices in this study. Figure 5b presents *IV* characteristics with a $V_g$ sweep from -15 V to 15 V. *IV* curves show non-linearity which can be attributed to a Schottky barrier at the contact interface. Additionally, the evolution of the transfer curve and FET performance metrics with increasing temperature can be found Figure S5.

We evaluate the performance of our representative WSe$_2$ phototransistor by comparing its mobility to that of other TMD-based devices reported in the literature, specifically those produced by other low-cost techniques such as EE, LPE, and selective-area solution deposition (SSD), as well as including the roll-to-roll MoS$_2$ device from our previous study[38]. As seen in Figure 5c, on average, LPE-based devices represent lower mobility values (< 0.1 cm$^2$ V$^{-1}$ s$^{-1}$), even though the vast majority of them were gated using ionic liquids. The LPE method offers advantages such as scalability and cost-effectiveness; however, reduced crystal quality poses challenges for integrating it into high-performance electronic device fabrication. In the case of electrochemical exfoliation, the reported mobility values distribute over a wide range, while some of them are lower than our representative device, there are considerable number of devices giving higher mobility values (> 10 cm² V⁻¹ s⁻¹). This is due to the ability of the method to finely tune the exfoliation parameters, such as changing the applied electrochemical potential or electrolyte type, thereby enabling control of the quality of the resulting product. Among WSe$_2$-based devices, to the best of our knowledge, only two studies have reported higher mobility values than ours; however, these results were achieved through the functionalization of the WSe$_2$ with molecular doping. Further details on the key device performance metrics reported in each study are summarized in Table S1 of (Supporting Information).

Figure 5d shows gate-controlled modulation of photocurrent cycles (left) and extracted $I_{pc}$ values as a function of the gate voltage (right). The measurement was carried out by sweeping the $V_g$ forward and backward between –10 V to 10 V in 1 V steps, during which time-resolved photocurrent cycles were recorded at each voltage level to track the evolution of the device's photoresponse under varying gate conditions. The corresponding results demonstrate that the amplitude of the photocurrent signal can be modulated by more than two orders of magnitude across the applied $V_g$ range. While the most positive $V_g$ (10 V) gives the highest $I_{pc}$ (~35 μA), the lower bound was obtained at -8 V (~0.22 μA), and beyond this $I_{pc}$ started to rise again with the emergence of hole conductance. Notably, photocurrent signals consistently retained a square-like shape throughout the entire $V_g$ sweep, indicating that the photoconduction effect is still the most dominant mechanism in photocurrent generation. While the photocurrent generation can be efficiently tuned by gate voltage modulation, the response time remained unchanged (< 20 ms), indicating that the photoresponse dynamics of the device are dominated by the intrinsic characteristics of the WSe$_2$ flakes and not by long-lived trap states. The stable and reversible switching behavior of the device under varying gate conditions demonstrates that roll-to-roll based WSe$_2$ phototransistors are suitable for applications requiring consistent and fast switching performance. To quantitatively evaluate the ability of the device for light detection, the responsivity ($R$) was calculated using the equation:

$$R = \frac{I_{pc}}{P_{light}} \times \frac{A_{spot}}{A_{effective}}$$

where $I_{pc}$ is the photocurrent, $P_{light}$ is the power of incident light, $A_{spot}$ is the area of the focused light spot, and $A_{effective}$ is the effective active channel area of the device. $A_{effective}$ was calculated by multiplying the channel area with the factor $c$. Figure 5e shows the variation of $I_{pc}$ and $R$ as a function of $P_{light}$ at a fixed wavelength of 625 nm. The photocurrent shows sub-linear behavior with increasing $P_{light}$ which can be attributed to the Auger recombination or the saturation of the trap states under strong light intensities that reduce the number of photogenerated charge carriers available to generate photocurrent[48,49]. Accordingly, $R$ shows an increasing trend with decreasing $P_{light}$, such that the maximum responsivity is extracted to be 117 A/W at minimum excitation intensity (~$3.1 \times 10^{-3}$ mW cm$^{-2}$). The plot showing the evolution of photocurrent cycles over time with varying $P_{light}$ can be found in Figure S6a. We further performed photocurrent measurements with varying wavelength to obtain the spectral sensitivity of the device. Figure S6b gives the $R$ as a function of incident light wavelength. The increased optical response around 750 nm is due to increased absorption originating from direct interband transitions in WSe$_2$. In order to investigate the spatial variations in photocurrent generation in the device channel, we monitored the photoactive regions through the scanning photocurrent measurement. The detailed discussion can be found in Figure S7.

Figure 5f presents the responsivity ($R$) as a function of response time to benchmark the performance of our device against previously reported WSe$_2$-based photodetectors. As shown, the responsivity of the roll-to-roll WSe$_2$ device exceeds that of all LPE-based devices and surpasses many of the devices fabricated via mechanical exfoliation (ME). Notably, in studies that report higher responsivity values than our device, performance enhancement was achieved through post-fabrication treatments (e.g., molecular doping, thermal annealing, chemical functionalization), or through device engineering approaches (eg. channel length scaling), or operation under strong gate bias conditions. In contrast, high responsivity from our roll-to-roll device originates intrinsically from the WSe$_2$ channel without any modification. Therefore, further external modification strategies need to be investigated to improve the performance of roll-to-roll based devices. Further details for each study represented in Figure 5f are given in Table S2 (Supporting Information).

Moreover, we continuously monitored the photocurrent response for 5000 s to reveal the stability of the device under long-term light exposure (see Figure S8). The results revealed that the current measured under illuminated and dark states shows a slight drift of 2.75% and 2.22%, respectively, indicating a relatively stable performance of the device under continuous light illumination.

**CONCLUSIONS**

In conclusion, this study demonstrates that integrating a controlled sliding motion into roll-to-roll mechanical exfoliation enables the scalable, uniform production of high-quality, wafer-scale 2D nanosheet films. AFM characterizations confirm that successive film transfers are essential for achieving well-covered substrates with highly interconnected flake networks. Additionally, the Beer-Lambert-based optical model effectively estimates spatial thickness variations across inch-sized films, showing good agreement with AFM measurements and highlighting its effectiveness for large-area thickness characterization. Statistical measurements of $WSe_2$ phototransistors reveal consistent electronic and optoelectronic performance, demonstrating the reproducibility of the method. Moreover, the fabricated devices exhibit high mobility and responsivity, showing competitive performance compared to those fabricated with electrochemical exfoliation. Overall, this work highlights the potential of roll-to-roll mechanical exfoliation for large-scale fabrication of high-performance devices based on 2D materials.

## MATERIALS AND METHODS

**Crystals and Substrates**

For the roll-to-roll exfoliation of the 2D crystals we used a transparent Nitto tape with the size of 30 mm × 100 mm (Nitto Denko Co. SPV224). Natural graphite flakes were supplied from ProGraphite GmbH. As a $MoS_2$ source, we used bulk natural molybdenite mineral (Tory Hill Mine, Quebec, Canada). Two different $WSe_2$ crystal sources were employed in this study: one synthesized in the laboratory and the other purchased commercially from HQ Graphene. $WSe_2$ was made by the CVT method in quartz ampoule by direct reaction from elements (selenium 2-6 mm granules, 99.9999%; tungsten, -100 mesh, 99.999%) using bromine as a vapor transport medium ($SeBr_4$, 99.9%). The stoichiometric amount of tungsten and selenium corresponding to 50 g of $WSe_2$ were placed in quartz ampoule (250 × 50 mm) together with 1 g of $SeBr_4$ and melt sealed under high vacuum ($< 1 \times 10^{-3}$ Pa, oil diffusion pump with liquid nitrogen cold trap) using oxygen-hydrogen welding torch. The ampoule was first placed in a muffle furnace and heated at 450 °C for 25 hours, subsequently at 500 °C for 50 hours, at 600 °C for 50 hours and at 800 °C for 50 hours. Heating and cooling rate were 1 °C min$^{-1}$. The ampoule with reacted $WSe_2$ was placed in a two zone furnace. First the growth zone was heated at 1000 °C and source zone at 700 °C. After 2 days the thermal gradient was reversed and the source zone was heated at 900 °C and growth zone at 800 °C for 12 days. Finally, the ampoule was cooled to room temperature and opened in an argon filled glovebox. h-BN was grown from iron metal flux (99.99%, 2-4 mm granules) using a boron nitride crucible in nitrogen-hydrogen atmosphere. The iron in the h-BN crucible was heated to 1600 °C and after 50 hours cooled to 1000 °C using a cooling rate of 10 °C/min and subsequently freely cooled to room temperature. Samples used for AFM measurements, large-area thickness estimation and large-batch photocurrent measurements were prepared using the in-lab synthesized $WSe_2$ crystal. The rest of the other samples were prepared using the commercially supplied $WSe_2$ crystal.

PC films with a thickness of 250 µm were purchased from Modulor (www.modulor.de). A wafer-scale PC substrate, used for transferring large-area $WSe_2$ films for thickness estimation, was prepared by cutting the full PC film using a benchtop smart cutting machine (Cricut Maker 3). To obtain the grid mask, extra-thick aluminum foil was patterned via a pulsed infrared laser (1064 nm) engraving system from Atomstack, Model M4[50]. Source-drain deposition masks were supplied from Ossila.

**Device Characterization**

Photocurrent measurements for large-batch $WSe_2$ devices were performed under ambient conditions using a home-built probe station. For the electronic transport measurements of large-batch $WSe_2$ phototransistors, we used Lake Shore CRX-6.5K cryogenic probe station, while single device measurements for complete electronic and optoelectronic characterizations were performed in a probe station embedded in a home-built vacuum chamber setup[51,52]. In both systems, measurements were carried out under vacuum conditions (~$10^{-5}$ mbar) after *in-situ* vacuum thermal annealing at 200 °C for 2 hours.

1) Electronic measurements:

    Two probes were connected to a Keithley 2450 source-meter unit to apply a bias voltage and monitor the current flow between the source and drain terminals. The third probe was connected to the silicon layer of the device to apply a back gate voltage using two programmable power supplies (TENMA 72-2715) connected in series.

2) Optoelectronic measurements:

For the light excitation, we used a fiber-coupled LED source with a wavelength of 625 nm (Thorlabs M625F2), which was controlled by a LED driver (Thorlabs LEDD1B). In power-dependent photocurrent measurements, the output light power from the LED source was modulated by altering the current driven through the LED driver with a TENMA unit. To reduce the power of light, absorptive neutral density filters were used (Thorlabs NEK01). Wavelength-dependent photocurrent measurements were performed using Bentham TLS120Xe tunable light source, which provides monochromatic illumination from 280 nm to 1100 nm.

**Atomic Force Microscopy**

A commercial Atomic Force Microscopy (AFM) system, from Nanotec, operating in ambient conditions was employed to perform morphological and thickness characterization of the samples. Measurements were acquired in dynamic mode with silicon tips (PPP-FMR from Nanosensors). Image analysis was conducted using Gwyddion software[53].

**Scanning Photocurrent Measurement**

Scanning photocurrent mapping was carried out under ambient conditions using a custom-built scanning photocurrent microscope[54]. The light excitation was provided by a 650 nm fiber-coupled laser (Optogear). The setup is based on a Motic microscope equipped with a 20× objective lens, providing a spot size resolution of approximately 1.5 µm. A Keithley 2450 source measurement unit (SMU) was used to monitor the device current under an applied source-drain bias of 5 V. The electrode channel was scanned using a motorized STANDA XY stage programmed with a 2 µm step size. Both the photocurrent signal and laser reflection were recorded simultaneously during the scan.

## ASSOCIATED CONTENT

**Supporting Information:** Supporting Information includes:

**Figure S1.** Roll-to-roll exfoliated large-area $WSe_2$ film on Nitto tape

**Figure S2.** $SiO_2/Si$ substrate coverage after single transfer of $WSe_2$ film

**Figure S3.** AFM images taken from $WSe_2$ films consist of different transfer numbers

**Figure S4.** Flake size distribution for $WSe_2$ nanosheets

**Figure S5.** Temperature-dependent transfer characteristics and corresponding FET performance metrics of a $WSe_2$ device

**Figure S6.** Power- and wavelength-dependent photoresponse in $WSe_2$

**Figure S7.** Scanning photocurrent measurement for $WSe_2$ device

**Figure S8.** Long-term durability test for $WSe_2$ device under light irradiation

**Table S1.** Comparison of mobility and $I_{on}/I_{off}$ values from transistors based on 2D materials fabricated with low-cost methods

**Table S2.** Comparison of photodetector performance metrics for $WSe_2$-based devices

## AUTHOR INFORMATION

**Corresponding Authors**

Yigit Sozen yigit.sozen@csic.es, Andres Castellanos-Gomez andres.castellanos@csic.es

**Author Contributions**

Y.S. lead data curation; formal analysis; investigation; methodology; the original draft writing. T.P., B.P.K., N.J.-A., J.H.R., Z.S., J.J.R., and C.M. supported data curation, formal analysis, investigation, and methodology. A.C.-G. lead conceptualization, funding acquisition, project administration, resources, supervision, supported the methodology, and original draft writing. All authors contributed equally to review and editing.

**ACKNOWLEDGEMENTS**

This work was funded by the Ministry of Science and Innovation (Spain) through the projects PRE2021-098348, PRE2022-105538, PID2023-151946OB-I00, and PDC2023-145920-I00 and funded by the European Commission – NextGenerationEU (Regulation EU 2020/2094), through CSIC's Quantum Technologies Platform (QTEP). The authors also acknowledge funding from the European Research Council (ERC) through the ERC-PoC 2024, StEnSo project (grant agreement no. 101185235), and ERC-2024 SyG SKIN2DTRONICS (grant agreement 101167218). N.J.A. acknowledges grant JDC2023-052025-I funded by MICIU/AEI/10.13039/501100011033 and by FSE+. J.J.R. hired under the Generation D initiative, promoted by Red.es, an organisation attached to the Ministry for Digital Transformation and the Civil Service, for the attraction and retention of talent through grants and training contracts, financed by the Recovery, Transformation and Resilience Plan through the European Union's Next Generation funds. Z.S. was supported by ERC-CZ program (project LL2101) from Ministry of Education Youth and Sports (MEYS) and by the project Advanced Functional Nanorobots (reg. No. CZ.02.1.01/0.0/0.0/15_003/0000444 financed by the EFRR). Z.S. also acknowledge the assistance provided by the Advanced Multiscale Materials for Key Enabling Technologies project, supported by the Ministry of Education, Youth, and Sports of the Czech Republic. Project No. CZ.02.01.01/00/22_008/0004558, Co-funded by the European Union.

## DATA AVAILABILITY

Data will be made available with request.

# Supporting Information:

# Wafer-Scale Films of Two-Dimensional Materials via Roll-to-Roll Mechanical Exfoliation

*Yigit Sozen[1]\*, Thomas Pucher[1], Bhagyanath Paliyottil Kesavan[2], Nuria Jiménez-Arévalo[1], Julia Hernandez-Ruiz[1], Zdenek Sofer[2], Carmen Munuera[1], Juan J. Riquelme[1], and Andres Castellanos-Gomez[1]\**

[1]*2D Foundry Research Group. Instituto de Ciencia de Materiales de Madrid (ICMM-CSIC), Madrid, E-28049, Spain.*

[2]*Department of Inorganic Chemistry, University of Chemistry and Technology Prague, Technická 5, 166 28 Prague 6, Czech Republic.*

\*corresponding authors: yigit.sozen@csic.es, andres.castellanos@csic.es

**Roll-to-roll exfoliated large-area $WSe_2$ film on Nitto tape**

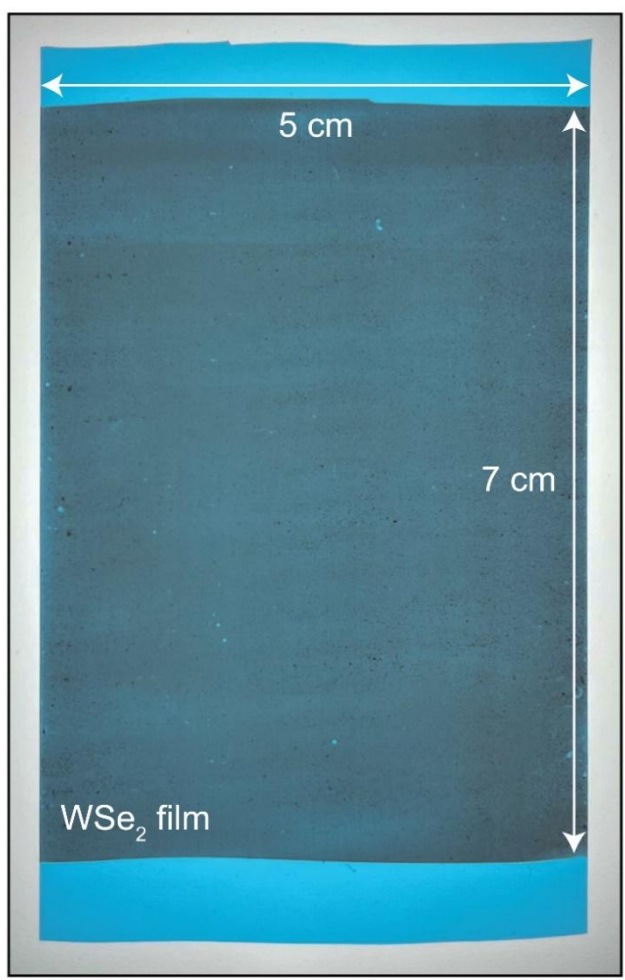

**Figure S 1.** Optical image of large area $WSe_2$ film on Nitto tape after roll-to-roll mechanical exfoliation.

**SiO$_2$/Si substrate coverage after single transfer of WSe$_2$ film**

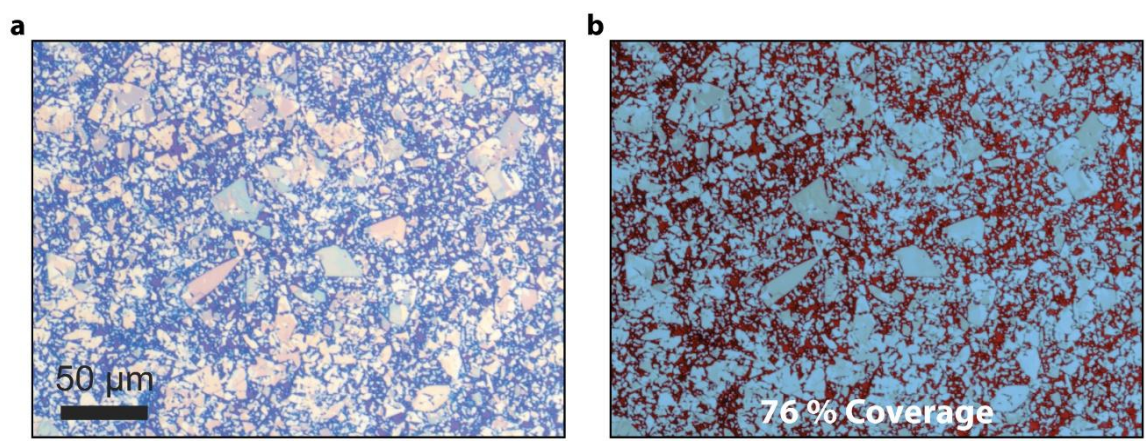

**Figure S2.** (a) Reflection mode optical micrograph of WSe$_2$ flakes after the transfer on SiO$_2$/Si substrate. (b) The masked image of optical micrograph highlights the flakes in blue, and substrate in red, indicating a 76 % substrate coverage. The coverage was obtained via Gwyddion software.

**AFM images taken from WSe$_2$ films consist of different transfer numbers**

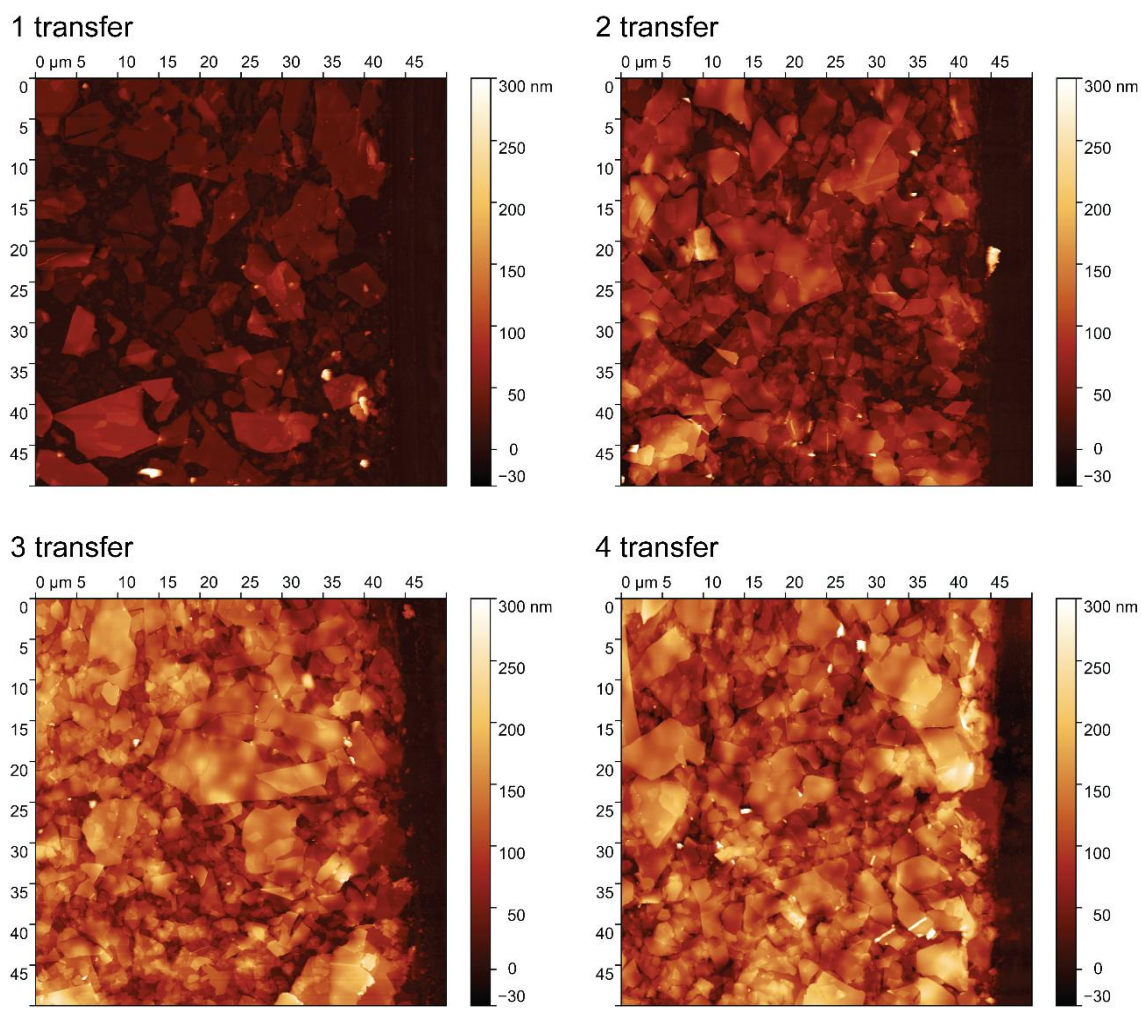

**Figure S3.** Full scan AFM images of WSe$_2$ films consist of 1, 2, 3, and 4 transfers.

# Flake size distribution for WSe$_2$ nanosheets

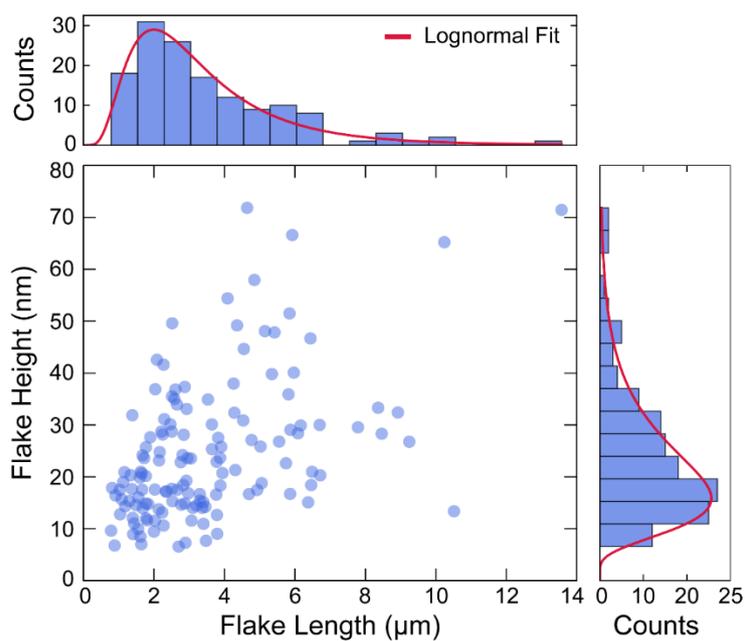

**Figure S4.** Flake length and height distribution of WSe$_2$ nanosheets.

# Temperature-dependent transfer characteristics and corresponding FET performance metrics of a WSe$_2$ device

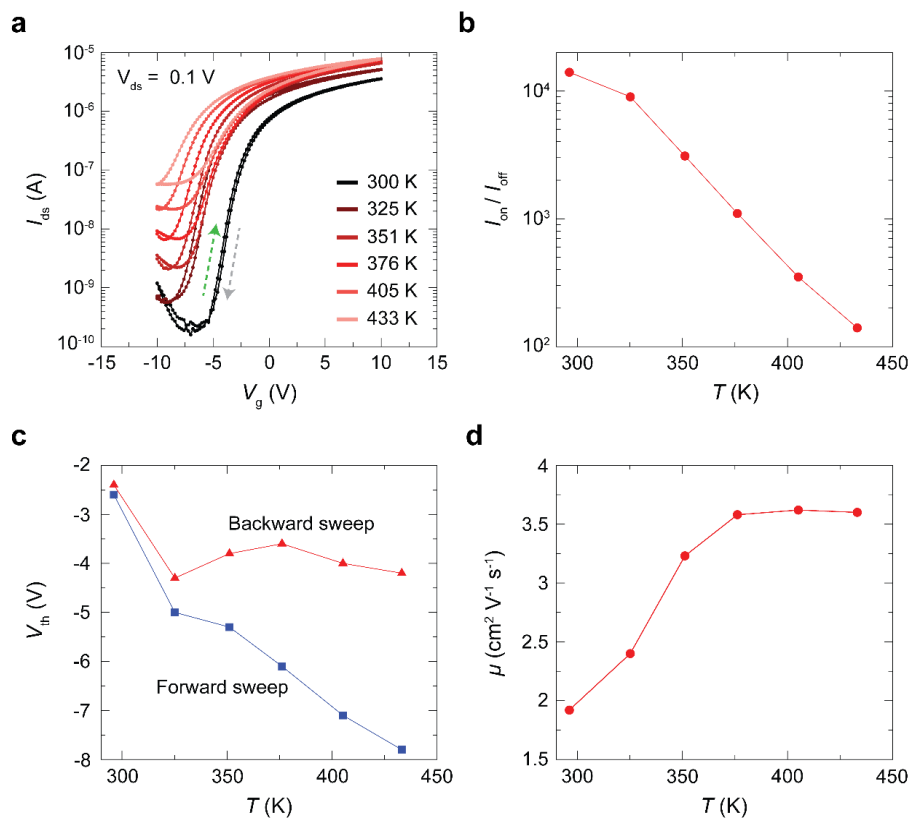

**Figure S5.** The evolution of the (a) transfer curve, (b) $I_{on}/I_{off}$ ratio, (c) $V_{th}$, and (d) $\mu$ with increasing temperature.

**Power- and wavelength-dependent photoresponse in WSe$_2$**

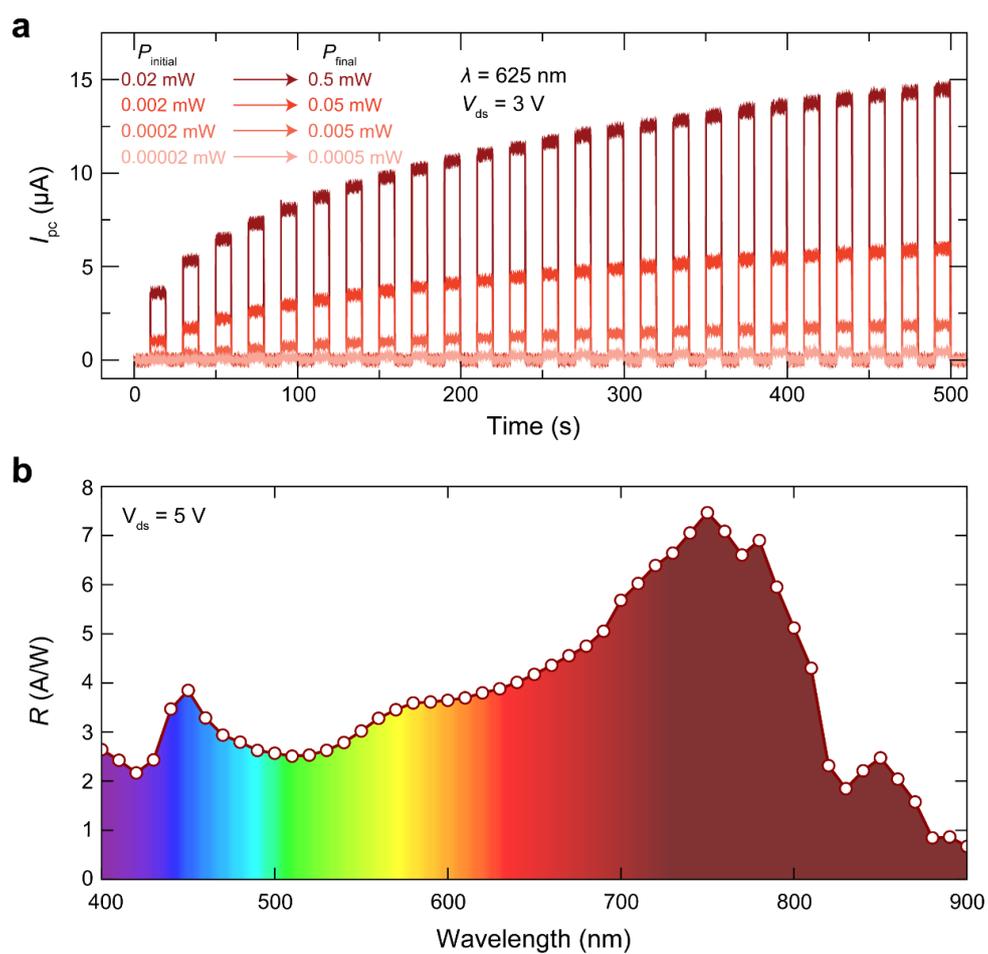

**Figure S6.** (a) Different sets of periodic ON/OFF cycles showing the evolution of photocurrent over increasing $P_{light}$. (b) $R$ as a function of excitation wavelength.

## Scanning photocurrent measurement for WSe$_2$ device

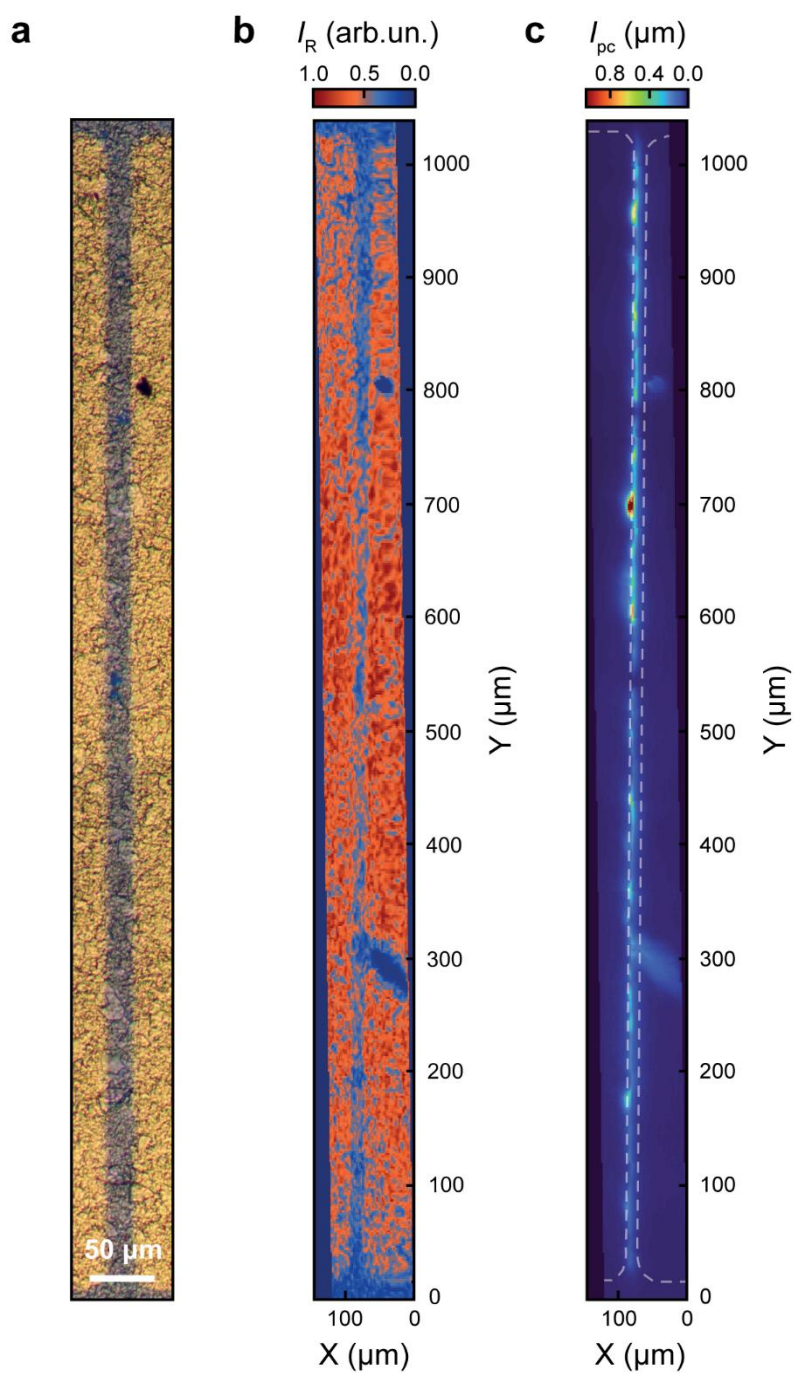

**Figure S7.** (a) Optical reflection microscope image under white light illumination, showing the transistor channel formed by 4 consecutive WSe$_2$ film transfers. Gold electrodes, defining a channel length of 30 μm, were deposited after the WSe$_2$ transfer. The film exhibits good overall homogeneity, with a small, uncovered region near the center of the channel. (b) Laser reflection intensity ($I_R$) map acquired using 650 nm illumination. Due to their metallic nature, the gold electrodes reflect more light and appear brighter, while the WSe$_2$ flakes in the channel show lower reflection intensity. Two dust particles are visible around 300 μm and 800 μm along the y-axis. (c) Corresponding scanning photocurrent map of the device measured under a 5 V drain-source bias. Bright areas indicate photocurrent generation, predominantly localized along the channel closer to the edge of one of the electrodes. While the entire channel contributes to photocurrent generation, the response is spatially inhomogeneous, with distinct 'hot spots' appearing around 600, 700, and 950 μm along the y-axis, suggesting a locally enhanced photoconductivity. A localized drop in photocurrent around 550 μm (y-axis) aligns with the uncovered area observed in (a). Dust particles seen in (b) also exhibit weak photocurrent, likely due to scattered light reaching adjacent regions.

**Long-term durability test for WSe$_2$ device under light irradiation**

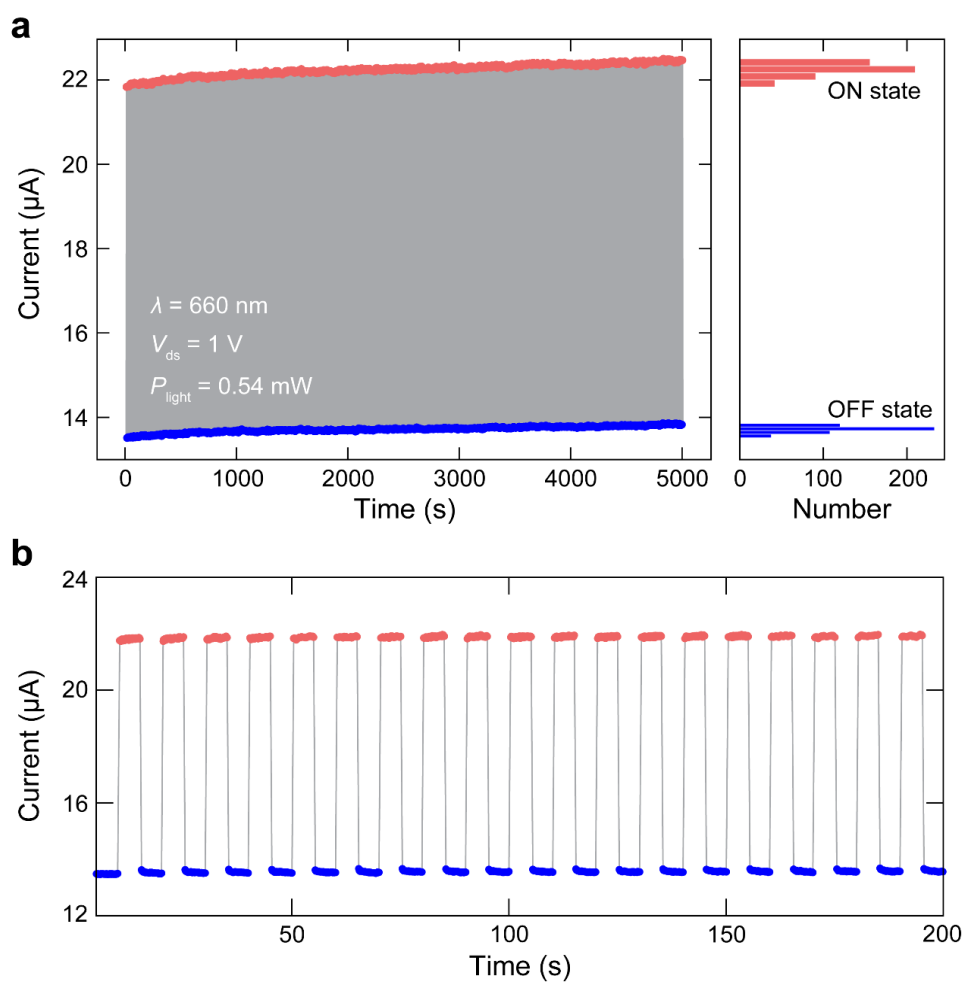

**Figure S8.** (a) Long-term durability test of the WSe$_2$ device under periodic illumination. The histogram plot on the right gives the distribution of the current values measured under ON and OFF conditions. The measurement was carried out for 5000 s by periodically exposing the device to the 660 nm LED light ($P_{light}$ = 0.54 mW) under a constant $V_{ds}$ (5 V), considering ON and OFF cycles with 5 s intervals. Data recorded at ON and OFF states are highlighted with red and blue, respectively. (b) A representative portion from the total set showing the initial 200 s to highlight the detailed photocurrent cycling behavior.

# Comparison of mobility and $I_{on}/I_{off}$ ratio values from transistors based on 2D materials fabricated with low-cost methods

| Material | Method | Carrier Type | Mobility (cm$^2$ V$^{-1}$ s$^{-1}$) | $I_{on}/I_{off}$ | Gating type | Reference |
|---|---|---|---|---|---|---|
| MoS$_2$ | EE | n-type | 11 | $2.6 \times 10^3$ | Ionic liquid | 1 |
| MoS$_2$ | EE | n-type | 0.5 | 3 | SiO$_2$ | 2 |
| MoS$_2$ | EE | n-type | 7-11 | $10^6$ | SiO$_2$ | 3 |
| MoS$_2$ | EE | n-type | 0.73 | $10^5$ | SiO$_2$ | 4 |
| MoS$_2$ | EE | n-type | 7 | $10^3$ | Ionic liquid | 5 |
| MoSe$_2$ | EE | n-type | 1.5 | $10^6$ | SiO$_2$ | 2 |
| WS$_2$ | EE | n-type | 9 | $3.4 \times 10^3$ | Ionic liquid | 1 |
| WS$_2$ | EE | n-type | 0.7 | $10^6$ | SiO$_2$ | 2 |
| WSe$_2$ | EE | Ambipolar | 1.3 (p) - 2 (n) | $4.2 \times 10^4$ | Ionic liquid | 1 |
| WSe$_2$ | EE | p-type | 1.5 | $10^6$ | SiO$_2$ | 6 |
| WSe$_2$ | EE | p-type | 27 | $10^7$ | HfO$_2$/SiO$_2$ | 7 |
| WSe$_2$ | EE | p-type | 35.22 | $10^5$ | SiO$_2$ | 8 |
| WSe$_2$ | EE | n-type | 0.002 | $10^2$ | SiO$_2$ | 2 |
| MoS$_2$ | LPE | n-type | 0.15 | ~$10^1$ | Ionic liquid | 9 |
| MoS$_2$ | LPE | n-type | 0.01 | $10^4$ | Ionic liquid | 10 |
| MoS$_2$ | LPE | n-type | 8.5 (0.12) | $10^5$ ($10^3$) | Al$_2$O$_3$ (SiO$_2$) | 11 |
| MoSe$_2$ | LPE | n-type | 0.18 | $10^2$ | Ionic liquid | 9 |
| WS$_2$ | LPE | p-type | 0.22 | $10^2$ | Ionic liquid | 9 |
| WS$_2$ | LPE | n-type | 0.01 | $10^4$ | Ionic liquid | 12 |
| WS$_2$ | LPE | n-type | 0.01 | $10^4$ | Ionic liquid | 10 |
| WS$_2$ | LPE | n-type | $6.2 \times 10^{-5}$ | 2.5 | SiO$_2$ | 13 |
| WSe$_2$ | LPE | p-type | 0.08 | $10^2$ | Ionic liquid | 9 |
| ReS$_2$ | LPE | n-type | 0.001 | $10^4$ | Ionic liquid | 10 |
| MoS$_2$ | SSD | n-type | 0.4 | $10^6$ | SiO$_2$ | 14 |
| MoS$_2$ | RR | n-type | 1.36 | ~3.5 | SiO$_2$ | 15 |
| WSe$_2$ | RR | Ambipolar | 3.4 | $10^4$ | SiO$_2$ | This work |

**Table S1.** Comparison of mobility values from 2D material-based transistors fabricated by different types of low-cost methods, such as electrochemical exfoliation (EE), liquid phase exfoliation (LPE) and selective-area solution deposition (SSD).

**Comparison of photodetector performance metrics for WSe$_2$-based devices**

| Method | Thickness (nm) | Wavelength (nm) | $R$ (A/W) | $V_{bias}$ (V) | $V_{gate}$ (V) | $\tau$ (ms) | Reference |
|---|---|---|---|---|---|---|---|
| ME | 54.2 | 650 | 7.55 – 25 | 0 – 3 | - | 0.33 | [16] |
| ME | 3-10 | 735 | 0.6 | 1 | 15 | 0.008 | [17] |
| ME | 39 | 500 | 0.17 | 1 | - | $8 \times 10^{-6}$ | [18] |
| ME | 2.3 | White light | 7 | 2 | -12 | 0.04 | [19] |
| ME | 4 | 532 | 5.16 | 3 | - | 0.533 | [20] |
| ME | Monolayer | 532 | 1027 | -1 | -3 | 0.4 | [21] |
| ME | 11 | 500 | $1.2 \times 10^5$ | 5 | - | < 2 | [22] |
| ME | 10–20 | 365 | $2.2 \times 10^6$ | -10 | - | 300 | [23] |
| ME | - | 520 | $1.27 \times 10^6$ | 1 | - | 2.8 | [24] |
| LPE | - | Pink | $3.65 \times 10^{-6}$ | 1 | - | 453 | [25] |
| LPE | - | 670 | $9.31 \times 10^{-5}$ | 0 | - | 130 | [26] |
| LPE | - | 470 | $3.31 \times 10^{-4}$ | 20 | - | $4.78 \times 10^3$ | [27] |
| LPE | 20 | AM1.5D | 1.5 | 1 | - | $< 1 \times 10^3$ | [28] |
| CVD | Monolayer | 650 | $1.8 \times 10^5$ | 2 | -60 | $5 \times 10^3$ | [29] |
| PLD | 48 | 635 | 0.92 | 10 | - | 900 | [30] |
| RR | 112 | 625 | 117 | 3 | - | < 20 | This work |

**Table S2.** Comparison of responsivity ($R$) and response time ($\tau$) values from WSe$_2$-based devices fabricated by different types of methods, such as mechanical exfoliation (ME), chemical vapor deposition (CVD), LPE, and pulsed-laser deposition (PLD).

## SUPPORTING INFORMATION REFERENCES